\documentclass[aip, reprint, amsmath,amssymb]{revtex4-1}

\usepackage{booktabs}
\usepackage{graphicx}
\usepackage{dcolumn}
\usepackage{bm}
\usepackage{color}
\usepackage{amsmath}
\usepackage{amssymb}
\usepackage{graphicx}
\usepackage[utf8]{inputenc}
\usepackage[T1]{fontenc}
\usepackage{mathptmx}
\usepackage{etoolbox}
\usepackage{mathptmx} 
\usepackage{url}
\usepackage{lineno}
\usepackage[colorlinks = true, linkcolor = blue, urlcolor  = blue, citecolor = blue, anchorcolor = blue]{hyperref}
\usepackage[textsize=footnotesize]{todonotes}

\usepackage{xspace}

\newcommand*{\bra}[1]{\ensuremath{\left<#1\right|}\xspace}%
\newcommand*{\ket}[1]{\ensuremath{\left|#1\right>}\xspace}%
\newcommand*{\braket}[2]{\ensuremath{\left<#1|#2\right>}\xspace}

\usepackage[textsize=footnotesize]{todonotes}
\usepackage{color}
\definecolor{oldtxtcolor}{rgb}{0.00, 0.0, 0.5}
\definecolor{newtxtcolor}{rgb}{0.00, 0.3867, 0.00}
\definecolor{newtxtcolor}{rgb}{0.00, 0.0, 1}
\definecolor{newtxtcolor2}{rgb}{1.00, 0.0, 0}
\definecolor{oldtxtcolor}{rgb}{1.00, 0.0, 0.00}

\def\verX{2}
\def\verO{1}
\def\verN{2}
\def\verON{12}

\ifx\verX\verO
 \newcommand { \oldtxt }[1] {{\color{oldtxtcolor}{#1}}}
 \newcommand { \newtxt }[1] {}
\fi
\ifx\verX\verN
 \newcommand { \oldtxt }[1] {}
 \newcommand { \newtxt }[1] {{\color{newtxtcolor}{#1}}}
\fi
\ifx\verX\verON
 \newcommand { \oldtxt }[1] {{\color{oldtxtcolor}{#1}}}
 \newcommand { \newtxt }[1] {{\color{newtxtcolor}{#1}}}
 
\fi

\begin{document}

\title{{Quantum tomography of molecules using ultrafast electron diffraction}}
\author{Jiayang Jiang$^*$}
\affiliation{Departments of Chemistry and Physics, University of Toronto, Toronto, Ontario M5S 3H6, Canada}
\author{Ming Zhang$^*$}
\email{zhming@stu.pku.edu.cn}
\affiliation{State Key Laboratory for Mesoscopic Physics and Frontiers Science Center for Nano-Optoelectronics, School of Physics, Peking University, Beijing 100871, China}
\author{Aosheng Gu}
\affiliation{Departments of Chemistry and Physics, University of Toronto, Toronto, Ontario M5S 3H6, Canada}
%
\author{R. J. Dwayne Miller}
\email{dmiller@lphys.chem.utoronto.ca}
\affiliation{Departments of Chemistry and Physics, University of Toronto, Toronto, Ontario M5S 3H6, Canada}
\author{Zheng Li}
\email{zheng.li@pku.edu.cn}
\affiliation{State Key Laboratory for Mesoscopic Physics and Frontiers Science Center for Nano-Optoelectronics, School of Physics, Peking University, Beijing 100871, China}
\affiliation{Collaborative Innovation Center of Extreme Optics, Shanxi University, Taiyuan, Shanxi 030006, China}
\affiliation{Peking University Yangtze Delta Institute of Optoelectronics, Nantong, Jiangsu 226010, China}


\begin{abstract}
%
%
We propose a quantum tomography (QT) approach to retrieve {the temporally evolving reduced density matrix in electronic state basis, where the populations and coherence between ground state and excited state are reconstructed from the ultrafast electron diffraction signal}.
In order to showcase the capability of the proposed QT approach, we simulate the nuclear wavepacket dynamics and ultrafast electron diffraction of photoexcited pyrrole molecules using ab initio quantum chemical CASSCF method.
{From} simulated time-resolved diffraction data, we retrieve the evolving density matrix in a crude diabatic representation basis and reveal the symmetry of the excited pyrrole wavepacket.
Our QT approach opens the route to make quantum version of ``molecular movie" that covers the electronic degree of freedom, and equips ultrafast electron diffraction with the power to reveal the {coherence between electronic states, relaxation and dynamics of population transfer}.
%
\end{abstract}

\maketitle
\section{Introduction}\label{sec:intro}

Quantum tomography (QT) plays an important role in quantum optics~\cite{Vogel1989:PRA40, Smithey1993:PRL70}, quantum information~\cite{Nielsen2002:0002-9505}, quantum computing~\cite{Torlai2018:NP14} and {molecular physics~\cite{Morrigan131:PRL23}}. By measuring multiple identical copies of the unknown quantum state systematically, the state and density matrix in the quantum system can be reconstructed, such as photonic states and states of matter waves~\cite{Molmer2003:PRL90}. 
Retrieving quantum states of molecules by QT would find important applications in ultrafast electron diffraction studies~\cite{Miller2014:Science343}. 
With time-resolved ultrafast electron diffraction patterns, one could investigate the structural changes in sub-picosecond time scales and connect to the changes in electron distributions~\cite{ischenko17:CR11066,gao13:343,Jiang2020:NC11,ishikawa15:1501,siwick03:302,Mu2023}. 
Based on the diffraction patterns, using QT to reconstruct the quantum states from the ultrafast electron diffraction patterns has been investigated for the rotational and vibrational wavepacket of molecules~\cite{zhang2021:NCOMM12}, and the dimension problem~\cite{Mouritzen2006:PRA73} in QT has been resolved by an iterative algorithm. 
{However, it remains a challenge to retrieve the temporal evolution of reduced density matrix elements in electronic state basis, which corresponds to the population and coherence between electronic states.}

\begin{figure}
        \includegraphics[width=0.5\textwidth]{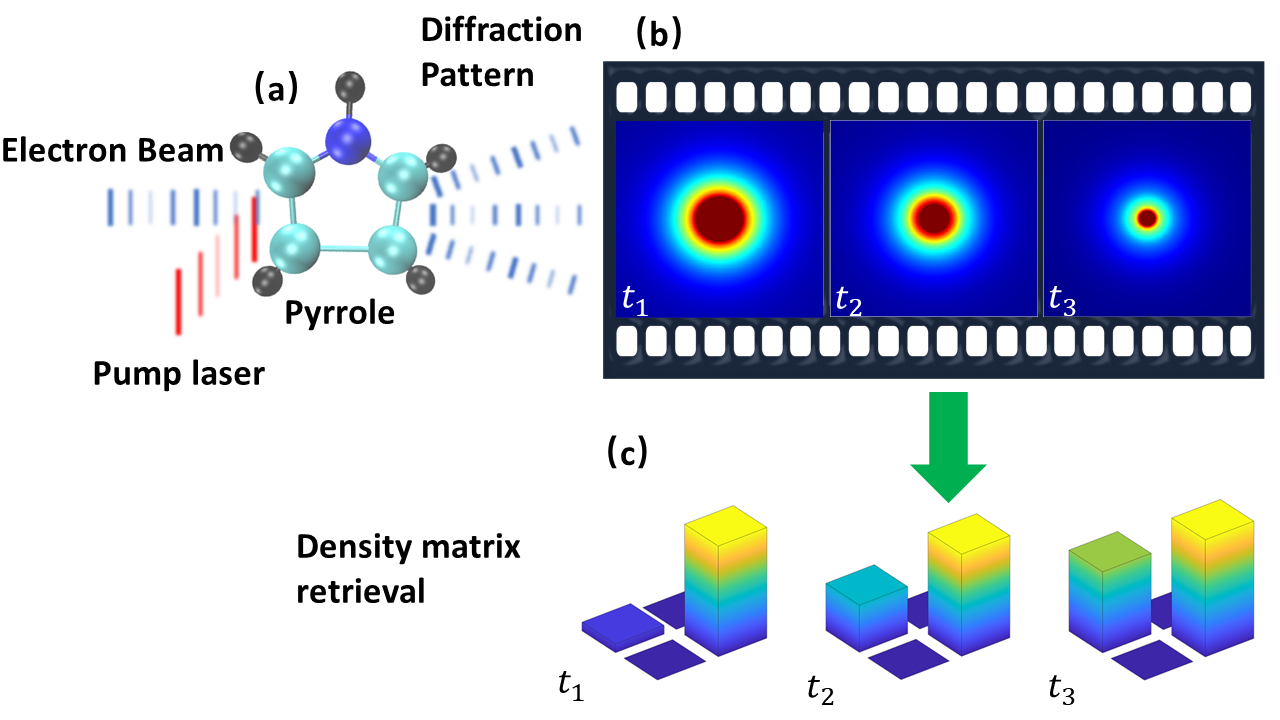}
\caption{ \label{fig:sche} Schematic of the quantum tomography (QT) from time-resolved ultrafast electron diffraction, illustrated by the photodissociation dynamics of pyrrole molecule.
(a) The photodissociation is triggered by the pump laser, and probed by the ultrafast electron pulse.
(b) By controlling the time delay between the pump laser and probe electron beam,  series of time-resolved diffraction images are generated for various time delays $t_i$. 
(c) Based on the proposed quantum tomography algorithms, the {reduced density matrices in electronic state basis} $\rho(t_i)$ are obtained from the diffraction patterns.
}
\end{figure}

In this work, we propose a quantum tomography (QT) approach to retrieve the evolving {reduced density matrix in electronic state basis} from ultrafast diffraction data.
The proposed QT method {reconstructs the temporal evolution of population and coherence between ground state and excited state} in crude diabatic representation, which is practically convenient for QT. 
To validate our QT approach, a simulated data set of ultrafast electron diffraction (UED) for the wavepacket dynamics of photoexcited pyrrole is generated by {ab initio quantum chemical  calculations~\cite{Robert19:JCTC1523,Northey2014:JCTC10,Carrascosa2017:PCCP19}} and analyzed by the QT algorithm. 
Our QT method recovers the {reduced density matrix in electronic state basis} for pyrrole molecule from the ultrafast electron diffraction data, as schematically illustrated in Fig.~\ref{fig:sche}.

In the following we briefly review the formulas for the nuclear wavepacket dynamics simulation of molecules, and the ab initio electron diffraction simulation.

\section{Theory}\label{sec:theory}
We take the pyrrole molecule as the test system. The relevant photochemical dynamics can be characterized by a model Hamiltonian involving two electronic states~\cite{Vallet2005:JCP123}.
The initial state after photoexcitation from the ground state $S_0$ could be $^1B_1$ and $^1A_2$ depending on the central wavelength of the excitation laser pulse, namely, 199~nm and 244~nm for $^1B_1$ and $^1A_2$ state respectively~\cite{Barbatti2006:JCP125}.
Both excited states relax through the conical intersection (CI) with the $S_0$ state via the coupled nuclear electronic dynamics.

In order to simulate the relaxation dynamics within interested reaction coordinates $\textbf{R}$ for the pyrrole system, we solve the time-dependent Schr\"{o}dinger equation of the nuclear wavepacket:

\begin{equation}\label{schro}
i\hbar \frac{\partial}{\partial t} \mathbf{\chi}(t) = \hat{H} \mathbf{\chi}(t) = \hat{T} \mathbf{\chi}(t) +\hat{V}\mathbf{\chi}(t)
\,,
\end{equation}
where $\mathbf{\chi}(t) = \begin{pmatrix} \chi_e(t) \\ \chi_g(t) \end{pmatrix}$ is the nuclear wavepacket in the two electronic states, i.e., $\chi_e(t), \chi_g(t)$ are the nuclear wavepacket in the ground state and excited state respectively, $\hat{T}$ and $\hat{V} = \begin{pmatrix} V_{ee}(\textbf{R}) & V_{eg}(\textbf{R}) \\ V_{ge}(\textbf{R}) & V_{gg}(\textbf{R}) \end{pmatrix} $ are the kinetic and diabatic potential energy operator, respectively.
We use the multi-configurational time-dependent Hatree (MCTDH) approach to simulate the wavepacket dynamics~\cite{Beck2000:PR324,MCTDH_8_5}. 
The wavepacket is equilibrated in the ground state, then the molecule is excited so the wavepacket is transferred from the ground state into the excited state upon photoexcitation. 
After the excitation, the wavepacket propagates in either $^1B_1-S_0$ or $^1A_2-S_0$ states.
From the evolving nuclear wavepacket $\chi_\alpha(t)$ of each electronic state, we can calculate the population probability, and the elements of the {reduced density matrix in electronic state basis} $\rho$:

\begin{equation}\label{eq:wv_nup}
\begin{split}
{\rho}_{\alpha \beta} &=  \braket{\chi_\alpha}{\chi_\beta} \\
&= \int \chi^*_\alpha(\textbf{R})\chi_\beta(\textbf{R})d\textbf{R}
\,,
\end{split}
\end{equation}
where {the diagonal and off-diagonal matrix elements are electronic state populations and coherences,} $\textbf{R}$ is the nuclear coordinate.

\subsection{Electronic wavefunction generation and electron diffraction dataset generation}

With calculated electronic wavefunctions $\Psi(\textbf{R})$ at a given reaction coordinate $\textbf{R}$ in different states, the scattering component between electronic states at scattering wavevector ${\textbf{s}}$ is:

\begin{equation}\label{eq:Ialpha_beta}
I_{\alpha,\beta}(\textbf{R},{\textbf{s}}) = \bra{\Psi_\alpha(\textbf{R})}\hat{L}_{\mathrm{e}}({\textbf{s}}) \hat{L}^{\dagger}_{\mathrm{e}}({\textbf{s}}) \ket{\Psi_\beta(\textbf{R})}
\,,
\end{equation}
where $\alpha,\beta$ refer to the electronic states.
{The diagonal term $I_{\alpha,\alpha}(\textbf{R},\textbf{s})$ is the scattering intensity from the $\alpha$-th electronic state, and the off-diagonal term $I_{\alpha,\beta}(\textbf{R},\textbf{s})$ is the coherent part of scattering intensity.}
The electron scattering operator $\hat{L}_{\mathrm{e}}$ is
\begin{equation}
\hat{L}_{\mathrm{e}}({\textbf{s}}) = \frac{1}{s^2}(\sum^{\mathrm{nuclei}}_J N_J e^{i{\textbf{s}} \cdot {\textbf{R}}_J} - \sum^{\mathrm{electrons}}_j e^{i{\textbf{s}} \cdot {\textbf{r}}_j})
\,.
\end{equation}
Here ${\textbf{s}}$ is the momentum transfer between incoming and outgoing electron in the diffraction, $N_J$ and ${\textbf{R}}_J$ are the charge and position of $J$-th nuclei, $\textbf{r}_j$ is the position of $j$-th electron. We can also identify the contribution of the total scattering. 
For a specific electronic state $\alpha$, the scattering intensity of elastic part $I_{\mathrm{elas}}$ and inelastic part $I_{\mathrm{inelas}}$ can be expressed as:

\begin{equation}\label{eq:elas_inelas}
\begin{split}
I_{\mathrm{elas}} &=  \left| \bra{\Psi_\alpha(\textbf{R})}\hat{L}_{\mathrm{e}}({\textbf{s}}) \ket{\Psi_\alpha(\textbf{R})} \right|^2 \,, \\ 
I_{\mathrm{inelas}} &= \sum_{j \neq \alpha} \bra{\Psi_\alpha(\textbf{R})}\hat{L}_{e}({\textbf{s}}) \ket{\Psi_j(\textbf{R})}\bra{\Psi_j(\textbf{R})} \hat{L}^{\dagger}_{e}({\textbf{s}}) \ket{\Psi_\alpha(\textbf{R})}\,.
\end{split}
\end{equation}

We simulate the ultrafast electron diffraction using the wavefunction from ab initio complete active space self-consistent field (CASSCF) calculations ~\cite{Northey2014:JCTC10,Carrascosa2017:PCCP19,Simmermacher2019:PRL122,Yang20:Sci885}.
The CASSCF wavefunction can be represented in the basis of configuration state function (CSF) $\ket{\Phi_{i}}$ as
\begin{equation}
\ket{\Psi}=\sum_i c_i \ket{\Phi_{i}}
    \,,
\end{equation}
{where $\ket{\Phi_{i}} = \frac{1}{\sqrt{N!}} (-1)^{\hat{P}} \hat{P} \Phi^\mathrm{H}_{i}$, $\hat{P}$ is the pairwise permutation operator and $\Phi^\mathrm{H}_{i} = u^1_{i}u^2_{i} \cdots u^N_{i}$ is the Hartree products of spin orbitals $u^j_{i}$. The spin orbitals $u^j_{i}$ are the products of spin functions $\kappa(j)$ and spatial molecular orbitals $\phi_j(\textbf{r}_j)$.} The electron diffraction intensity at momentum transfer $\textbf{s}$ can be further calculated via the reduced one- and two-electron reduced density operators as:

\begin{equation}\label{eq:I_rho1_rho2}
\begin{split}
I(\textbf{s}) =& \dfrac{1}{s^{4}} [ \sum_{I,J} N_{I}N_{J} e^{i \textbf{s} \cdot (\textbf{R}_{I} - \textbf{R}_{J})} \\ &- 2\sum_{I}N_{I}\int e^{i \textbf{s} \cdot (\textbf{R}_{I} - \textbf{r})}\rho (\textbf{r}) d\textbf{r} \\ 
&+ n + \int e^{i \textbf{s} \cdot (\textbf{r} - \textbf{r}')}\rho(\textbf{r},\textbf{r}')d\textbf{r}d\textbf{r}' ]
\,,    
\end{split}
\end{equation}
where $n$ is the total number of electrons in the molecule, $\textbf{N}_{I}$, $\textbf{N}_{J}$ are the charge of $I$-th, $J$-th nuclei respectively, $\textbf{R}_{I}$ is the position of $I$-th atom. $\rho (\textbf{r})$ and $\rho(\textbf{r},\textbf{r}')$ are the reduced one and two electron density operators. {In the molecular orbital basis, $\rho (\textbf{r}) = \sum_{i,j} c_{i,j} \phi_i(\textbf{r}) \phi_j(\textbf{r})$ and $\rho(\textbf{r},\textbf{r}') = \sum_{i,j,k,l} d_{i,j,k,l} \phi_i(\textbf{r}) \phi_j(\textbf{r}) \phi_k(\textbf{r}') \phi_l(\textbf{r}')$ are the linear combinations of spatial molecular orbitals products, and spatial molecular orbitals are expanded by Gaussian-type orbitals (GTOs). 
So the Fourier transform in Eq.~\ref{eq:I_rho1_rho2} are linear combinations for Fourier transform of GTOs products, which could be calculated analytically~\cite{Northey2014:JCTC10}.}

We can calculate the rotational average of the diffraction pattern for molecules in both isotropic distributions and laser-aligned anisotropic distributions. {To calculate the rotational average precisely and quickly, a proper distribution of grid points should be established, and Lebedev quadrature satisfy the requirements. In Lebedev quadrature, quadratures are invariant with octahedron rotation group with inversion $G^*_8$~\cite{LEBEDEV197610}. Lebedev quadrature uses fewer grid points, but the integration accuracy is comparable with other quadratures}. We use Lebedev grid points either in isotropic distributions or in anisotropic distributions~\cite{Robert19:JCTC1523}. 

%

After calculating the expectation of the scattering operators in Eq.~\ref{eq:Ialpha_beta}, the total scattering intensity for the nuclear wavepacket is calculated as:
{
\begin{equation}\label{eq:Itotal}
I({\textbf{s}},t) = \sum_{\alpha,\beta} \int d\textbf{R}{\chi^*_\alpha(\textbf{R},t)}I_{\alpha,\beta}(\textbf{R},{\textbf{s}}){\chi_\beta (\textbf{R},t)}
\,,
\end{equation}}
for the nuclear wavepacket $\chi(\textbf{R},t)$.
{Eq.~\ref{eq:Itotal} is calculated numerically by integrating the scattering matrix component $I_{\alpha,\beta}(\textbf{R},{\textbf{s}})$ over the nuclear degrees of freedom using the simulated nuclear wavepackets.
The information of reduced density matrix in electronic state basis, including the temporal evolution of populations and coherence, are encoded in the diffraction signal~\cite{Kowalewski2017:SD4,Keefer2021:PNAS118,PhysRevLett.129.103001,Simmermacher2019:PRL122}. }

In the real electron diffraction experiment, the temporal resolution is determined by the pulse width of the electron beam. Suppose the incident electron beam is a Gaussian beam, the scattering intensity is
\begin{equation}\label{eq:Irefine}
I_{\mathrm{refine}}({\textbf{s}},t) = \dfrac{1}{\sqrt{2\pi}\sigma_{\mathrm{t}}} \int I({\textbf{s}},t') e^{-\dfrac{(t-t')^2}{2\sigma_{\mathrm{t}}}} dt'
\,,
\end{equation}
where $I({\textbf{s}},t')$ is the scattering intensity calculated in Eq.~\ref{eq:Itotal}, {and the full width at half maximum (FWHM) pulse width is $2\sqrt{2\sigma_\mathrm{t}\ln{2}}$}.
\begin{figure}
        \includegraphics[width=0.5\textwidth]{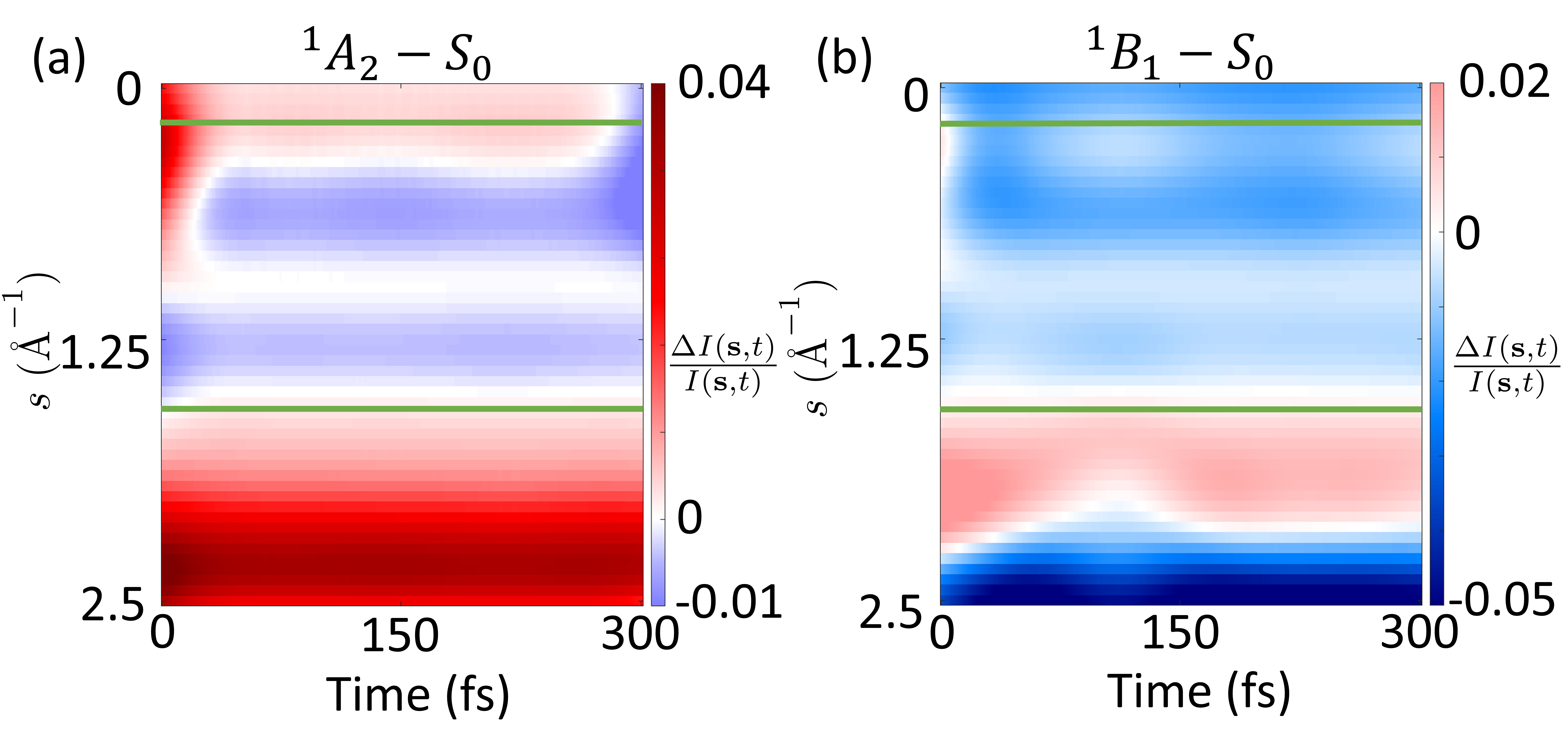}
 \caption{ \label{fig:py_approx} {The relative difference of total scattering intensity $\frac{\Delta I(\textbf{s},t)}{I(\textbf{s},t)}$ defined in Eq.~\ref{Eq:rel_diff} for (a) the ${}^{1}A_{2}-S_0$ states and (b) the ${}^{1}B_{1}-S_0$ states respectively}, which reflects the change of diffraction intensity due to nuclear motion. {The data between the two green lines in each figure is used for quantum tomography.}
The FWHM of {the electron pulse} is 50~fs. The time range is from 0~fs to 300~fs.
%
}
\end{figure}

\begin{figure}
        \includegraphics[width=8.3cm]{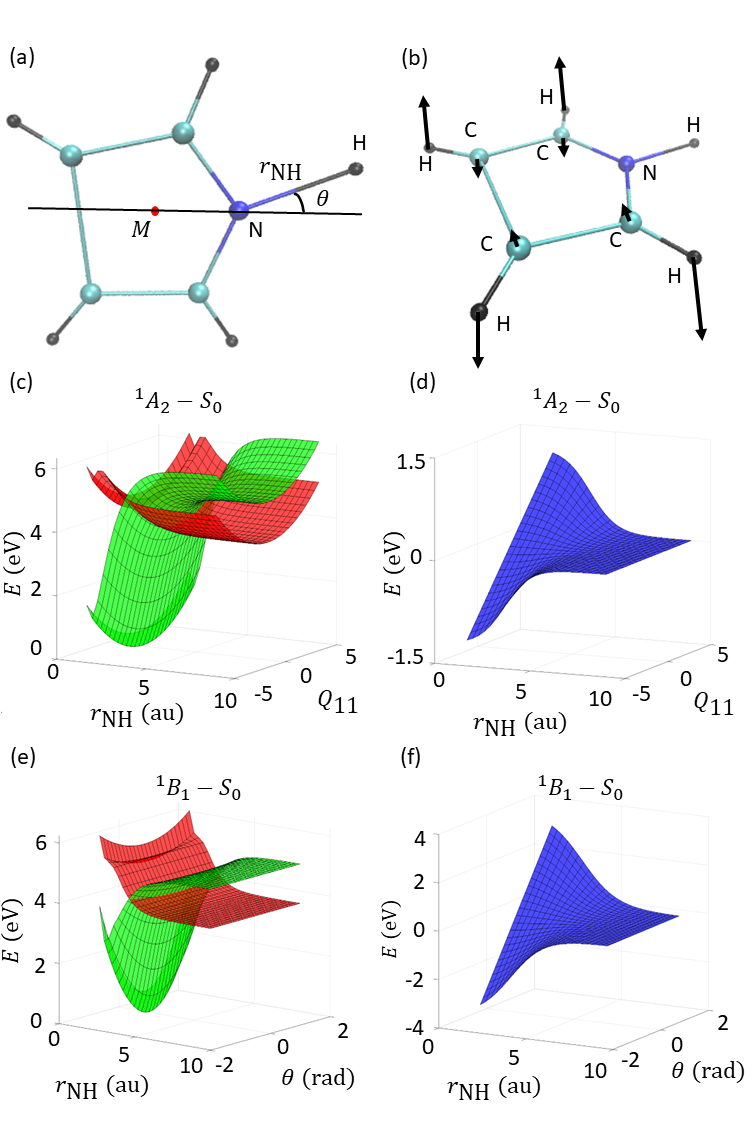}
\caption{ \label{fig:py_hamil} {Reaction coordinates and diabatic Hamiltonian matrix elements of the pyrrole molecule.}
(a) The reaction coordinate $r_\mathrm{NH}$ of ${}^{1}B_{1}$ and ${}^{1}A_{2}$ states and the coupling mode $\theta$ of the ${}^{1}B_{1}-S_0$ states.
(b) The nuclear displacement vectors of the coupling mode $Q_{11}$ of the ${}^{1}A_{2}-S_0$ states.
(c) The diabatic potential energy surface (PES) and (d) off-diagonal Hamiltonian matrix elements for ${}^{1}A_{2}-S_{0}$ states, the corresponding coupling mode is $Q_{11}$.
(e) The diabatic PES and (f) off-diagonal Hamiltonian matrix elements for ${}^{1}B_{1}-S_{0}$ states, the corresponding coupling mode is $\theta$.
}
\end{figure}

\subsection{Electronic state density reconstruction}\label{sec:reconstruct}
%
%
{From Eq.~\ref{eq:Itotal}, the information of $\rho_{\alpha\beta}(t)= \int \chi^{*}_\alpha(\textbf{R},t) \chi_\beta (\textbf{R},t) d\textbf{R}$ is encoded in electron diffraction intensity.
However, the {reduced density matrix in electronic state basis} can not be directly retrieved from diffraction intensity, due to the $\textbf{R}$ dependence of $I_{\alpha,\beta}(\textbf{R},\textbf{s})$ defined in $\Psi_\alpha(\textbf{R})$ diabatic representation basis as Eq.~\ref{eq:Ialpha_beta}.
To reconstruct the {reduced density matrix elements $\rho_{\alpha\beta}(t)$} from ultrafast electron diffraction data, we use 
\begin{equation}\label{eq:Ialpha_beta_R0}
I_{\alpha,\beta}(\textbf{R}_0,{\textbf{s}}) = \bra{\Psi_\alpha(\textbf{R}_0)}\hat{L}_\mathrm{e}({\textbf{s}}) \hat{L}^{\dagger}_\mathrm{e}({\textbf{s}}) \ket{\Psi_\beta(\textbf{R}_0)}\,,
\end{equation}
at a fixed nuclear coordinate $\textbf{R}_0$ to approximate $I_{\alpha,\beta}(\textbf{R},\textbf{s})$, then the total diffraction intensity is approximated by
\begin{equation}\label{eq:Itotal_app}
\begin{split}
\tilde{I}({\textbf{s}},t) &= \sum_{\alpha,\beta} \int d\textbf{R}{\chi^*_\alpha(\textbf{R},t)}{\chi_\beta (\textbf{R},t)}I_{\alpha,\beta}(\textbf{R}_0,{\textbf{s}}) \\
    &= \sum_{\alpha,\beta} \rho_{\alpha\beta}(t) I_{\alpha,\beta}(\textbf{R}_0,{\textbf{s}})
    \,.
\end{split}
\end{equation}
The approximation of using $I_{\alpha,\beta}(\textbf{R}_0,{\textbf{s}})$ defined in Eq.~\ref{eq:Ialpha_beta_R0} to replace $I_{\alpha,\beta}(\textbf{R},{\textbf{s}})$ defined in Eq.~\ref{eq:Ialpha_beta}, is equivalent to use the crude diabatic representation of electronic state $\ket{\Psi_\alpha (\textbf{R}_0)}$ to replace another diabatic representation $\ket{\Psi_\alpha (\textbf{R})}$ used in MCTDH wavepacket simulations.}
{In crude diabatic representation, the diabatic electronic state basis $\ket{\Psi_\alpha (\textbf{R}_0)}$ is taken to be the electron eigenstate wavefunctions of the equilibrium nuclear coordinates $\textbf{R}_0$ in the ground state~\cite{Zhang1998:WS_Theory}. 
The definition of crude diabatic representation is rigorous without approximation, but the density matrix in crude diabatic representation is different from the density matrix in another diabatic representation used in MCTDH calculation.
The small difference for density matrices $\rho_{\alpha\beta}(t)$ in these two representations, which are shown below in Fig.~\ref{fig:recons} due to the small amplitude of nuclear motion for photoexcited pyrrole molecules, demonstrates the advantage and feasibility of using the crude diabatic representation.
}
%
{Although the electronic state basis chosen for QT is different from that used in MCTDH wavepacket dynamics calculation, the density matrices $\rho_{\alpha\beta}(t)$ in these two diabatic representations exhibit similar characters within small differences, as shown in Fig.~\ref{fig:recons} below.
%

%
%
%
\begin{figure}
        \includegraphics[width=0.5\textwidth]{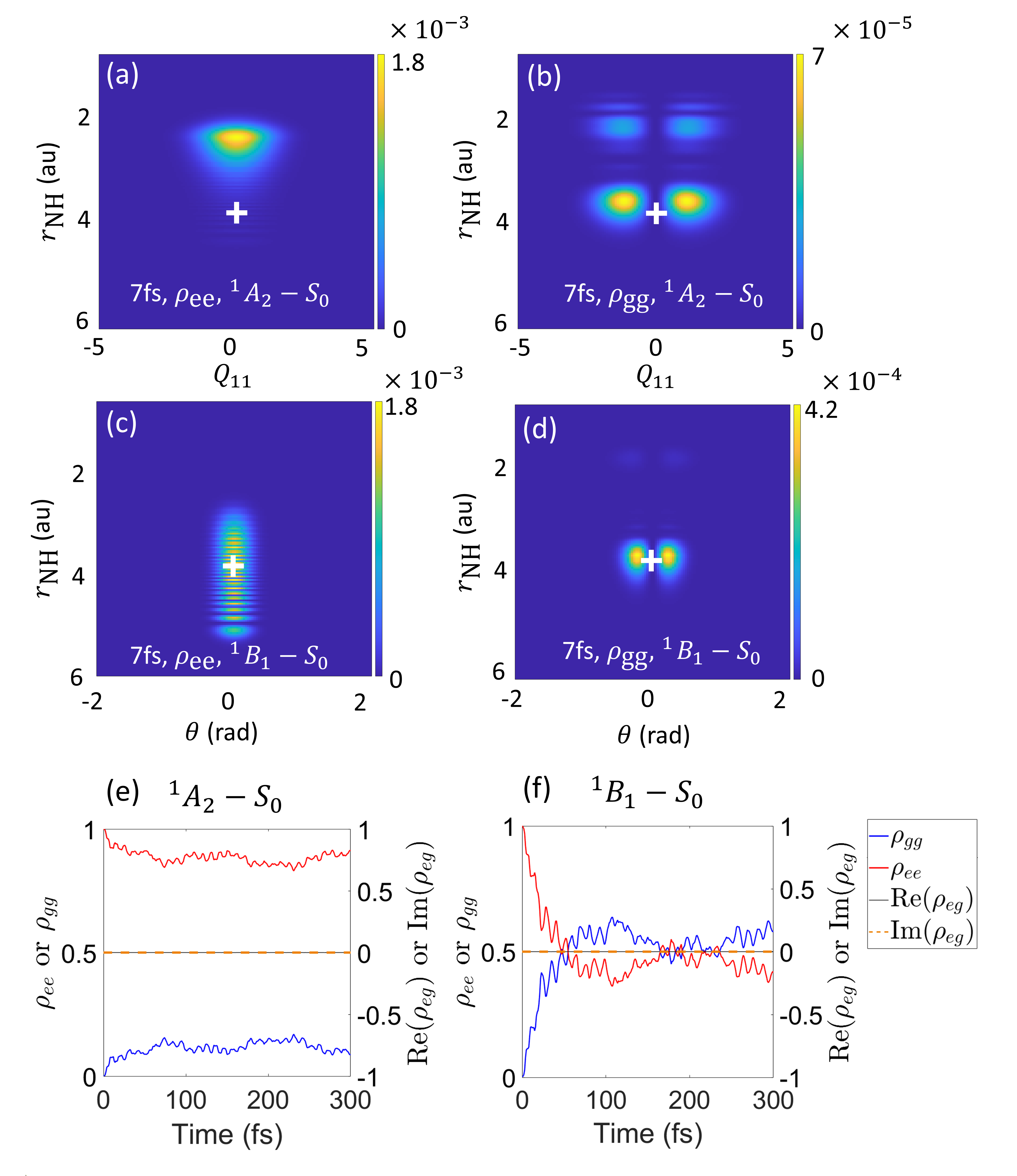}
{\caption{\label{fig:mctdh} The nuclear wavepacket dynamics of pyrrole. The snapshots of the nuclear wavepacket at 7~fs for the pyrrole wavepacket dynamics (a)(b) for the ${}^{1}A_{2}-S_0$ states and (c)(d) for the ${}^{1}B_{1}-S_0$ states, respectively. The white crosses in (a)-(d) indicate the conical intersection.
(a)(b) are the probability density snapshots in the excited state ${}^{1}A_{2}$ and ground state $S_0$ respectively. The wavepacket density are given as function of reaction coordinates $r_\mathrm{NH}$ in atomic unit (au) and $Q_{11}$. 
Panels (c) and (d) are the probability density of nuclear wavefunction snapshots in the excited state ${}^{1}B_{1}$ and ground state $S_0$ respectively, where $r_{\mathrm{NH}}$ and $\theta$ are the reaction coordinates during the wavepacket dynamics. 
Panels (e) and (f) are the temporally evolving reduced density matrix elements for the  ${}^{1}A_{2}-S_0$ and ${}^{1}B_{1}-S_0$ state, respectively.
The diagonal elements $\rho_\mathrm{ee}$ and $\rho_\mathrm{gg}$ are the populations in the excited state and the ground state, respectively. $\rho_\mathrm{eg}$ is the coherence between ground state and excited state. The wavepacket is initially in the ${}^{1}A_{2}$ and ${}^{1}B_{1}$ excited states.}
}
\end{figure}

To illustrate the accuracy of the approximation $I_{\alpha,\beta}(\textbf{R},{\textbf{s}})\approx I_{\alpha,\beta}(\textbf{R}_0,{\textbf{s}})$, we calculate the relative difference of diffraction intensity defined by
\begin{equation}\label{Eq:rel_diff}
    \frac{\Delta I(\textbf{s},t)}{I(\textbf{s},t)}=\frac{I(\textbf{s},t)-\tilde{I}({\textbf{s}},t)}{I(\textbf{s},t)}\,,
\end{equation}
where $I(\textbf{s},t)$ and $\tilde{I}(\textbf{s},t)$ are defined in Eq.~\ref{eq:Itotal} and Eq.~\ref{eq:Itotal_app}, respectively.
As shown in Fig.~\ref{fig:py_approx}, the relative change of diffraction intensity caused by nuclear motion for pyrrole molecule is {approximately} $0.5\%$ {within a specific region of momentum transfer}, which validates the approximation $I_{\alpha,\beta}(\textbf{R},{\textbf{s}})\approx I_{\alpha,\beta}(\textbf{R}_0,{\textbf{s}})$.
Also, it demonstrates that the crude diabatic basis $\ket{\Psi(\textbf{R}_0)}$ for QT retrieval algorithm, is close to the diabatic basis $\ket{\Psi(\textbf{R})}$ used in MCTDH wavepacket dynamics.

However, when generalizing this QT retrieval method to other molecules, the effect of nuclear motion to the change of diffraction intensity should be checked.
The validity of approximation $I_{\alpha,\beta}(\textbf{R},{\textbf{s}})\approx I_{\alpha,\beta}(\textbf{R}_0,{\textbf{s}})$ holds when the amplitude of atomic motion is small.
When the approximation does not hold, such as in molecular dissociation processes, the QT algorithm can be enhanced by replacing the fixed nuclear geometry $\textbf{R}_0$ in crude diabatic basis $\ket{\Psi(\textbf{R}_0)}$ by the time-dependent nuclear geometry $\textbf{R}_0(t)$, which is estimated by ab initio calculation or retrieved from diffraction intensity.
}

For pyrrole system, we can write Eq.~\ref{eq:Itotal_app} in the matrix form:
\begin{equation}
\label{eq:total_mat}
\begin{split}
 \begin{pmatrix} I({\textbf{s}_1}) \\
 I({\textbf{s}_2}) \\  \vdots \\ I({\textbf{s}_n}) \end{pmatrix} &= I(\textbf{R}_0) \begin{pmatrix} \rho_\mathrm{gg} \\ \rho_\mathrm{ge} \\  \rho_\mathrm{eg} \\ \rho_\mathrm{ee} \end{pmatrix}\,, \\ 
  I(\textbf{R}_0) &= \begin{pmatrix}I_\mathrm{gg}(\textbf{R}_0,{\textbf{s}_1}) & I_\mathrm{ge}(\textbf{R}_0,{\textbf{s}_1}) & I_\mathrm{eg}(\textbf{R}_0,{\textbf{s}_1}) & I_\mathrm{ee}(\textbf{R}_0,{\textbf{s}_1}) \\ I_\mathrm{gg}(\textbf{R}_0,{\textbf{s}_2}) & I_\mathrm{ge}(\textbf{R}_0,{\textbf{s}_2}) & I_\mathrm{eg}(\textbf{R}_0,{\textbf{s}_2}) & I_\mathrm{ee}(\textbf{R}_0,{\textbf{s}_2}) \\ \vdots & \vdots & \vdots & \vdots \\ I_\mathrm{gg}(\textbf{R}_0,\textbf{s}_n) & I_\mathrm{ge}(\textbf{R}_0,\textbf{s}_n) & I_\mathrm{eg}(\textbf{R}_0,\textbf{s}_n) & I_\mathrm{ee}(\textbf{R}_0,\textbf{s}_n) \end{pmatrix}\,, \\ 
   \rho_\mathrm{ge} &= \rho^{*}_\mathrm{eg}\,, \\  %
   \rho_\mathrm{ee} &+ \rho_\mathrm{gg} = 1\,.
\end{split}
\end{equation}
With the simulated electron diffraction intensity, we can retrieve the density matrix in the diabatic representation based on Eq.~\ref{eq:total_mat}.
Since the number of variables for the density matrix is smaller than the diffraction intensity data, retrieving the density matrix by Eq.~\ref{eq:total_mat} is an over-determined linear system; we use the least square method to recover the complex density matrix $\rho_{\mathrm{rec}}$ with elements $\rho_{\alpha\beta}$ at time $t$.
\begin{equation} 
\label{eq:rho_min_retrieve}
\rho_{\mathrm{rec}} = \underset{\rho}{\mathrm{arg}\,\mathrm{min}} \{ ||  I(\textbf{R}_0)\rho - I_{\mathrm{refine}}(t) ||_2 ^ 2 \} \,, 
\end{equation}
where $\rho = \begin{pmatrix} \rho_\mathrm{gg} \\ \rho_\mathrm{ge} \\  \rho_\mathrm{eg} \\ \rho_\mathrm{ee} \end{pmatrix}$ \,, 
$I_{\mathrm{refine}}(t) =  \begin{pmatrix} I_{\mathrm{refine}}({\textbf{s}_1,t}) \\
 I_{\mathrm{refine}}({\textbf{s}_2},t) \\  \vdots \\ I_{\mathrm{refine}}({\textbf{s}_n},t) \end{pmatrix}$ \, and
 $I_{\mathrm{refine}}({\textbf{s}_i,t})$ is calculated in Eq.~\ref{eq:Irefine}.
Given diffraction data at each time delay $t_i$, the {temporally evolving reduced density matrix in electronic state basis} can be recovered.

\subsection{Model Hamiltonian of photoexcited pyrrole molecule}\label{sec:Hmodel}

We simulate the wavepacket dynamics of photoexcited pyrrole molecule for ${}^{1}B_{1}-S_0$ and ${}^{1}A_{2}-S_0$ states.
For the dynamics from the initial state ${}^{1}B_{1}$, pyrrole can be qualitatively regarded to contain 3 parts: active hydrogen, nitrogen, and remaining atoms $M$. 
$r_\mathrm{NH}$ and $\theta$ are the reaction coordinates, as shown in Fig.~\ref{fig:py_hamil}(a). 
$r_\mathrm{NH}$ is the length of the N-H bond and $\theta$ is the angle between $\textbf{r}_\mathrm{MN}$ and $\textbf{r}_\mathrm{NH}$. $\textbf{r}_\mathrm{MN}$ is the displacement vector pointing from the position of the mass center of $M$ to the position of the nitrogen atom. Here, $\textbf{r}_\mathrm{NH}$ is the displacement vector pointing from the nitrogen atom to the active hydrogen atom. 
For the dynamics in the ${}^{1}A_{2}-S_0$ states, $r_\mathrm{NH}$ and $Q_{11}$ are the major reaction coordinates, where $Q_{11}$ is the nuclear displacement coupling modes illustrated in Fig.~\ref{fig:py_hamil}(b). %
The potential energy surface (PES) between ${}^{1}A_{2}-S_0$, ${}^{1}B_{1}-S_0$ conical intersection are fitted by Vallet et al. ~\cite{Vallet2004:FD127,Vallet2005:JCP123}.
Two reaction coordinates, the N-H distance $r_\mathrm{NH}$ and the coupling mode between different potential surfaces, are presented in Fig.~\ref{fig:py_hamil}(c)-(f).
{We perform two separate MCTDH wavepacket dynamics simulations of photoexcited pyrrole molecule, both including two electronic states.
For the ${}^{1}B_{1}-S_0$ wavepacket dynamics, the coordinates $r_\mathrm{NH}$ and $\theta$ are included.
And for the ${}^{1}A_{2}-S_0$ wavepacket dynamics, the coordinates $r_\mathrm{NH}$ and $Q_{11}$ are included.}

\section{Results}\label{sec:res}
We apply the quantum tomography approach to analyze the electronic dynamics of the photoexcited pyrrole molecule.
It is found that the non-adiabatic couplings around conical intersections (CI) regions play important roles in the dynamical process of pyrrole~\cite{Blank1994:CP187,Wei2004:FD127,Vallet2005:JCP123}.
{Transient electronic coherence can be created when the molecular wavepacket passes through the CI.}
The population dynamics and coherence can be revealed quantitative by the QT approach.

\subsection{Analysis of nuclear wavepacket}
We simulate the coupled electronic nuclear wavepacket dynamics in the ${}^{1}B_{1}-S_0$ and ${}^{1}A_{2}-S_0$ states of photoexcited pyrrole molecule.
In the calculations, the grid point number in N-H stretching mode is 256, ranging from 1 to 9 atomic unit (au). For the coupling mode, the grid point number is 128, ranging from -2 to 2~rad in ${}^{1}B_{1}-S_0$ and {-5 to 5} in ${}^{1}A_{2}-S_0$. The nuclear wavepackets are propagated to 300~fs, and the time step is 0.05~fs. 
In the simulations, the pyrrole molecule is excited from the ground state $S_0$ to the excited state at 0~fs.
The {reduced density matrix in electronic state basis} can be obtained from the nuclear wavepacket $\chi(t)$ at time $t$ as in Eq.~\ref{eq:wv_nup}. 
The snapshots of nuclear wavepacket and reduced density matrix elements are shown in Fig.~\ref{fig:mctdh}, we can see that the population of the excited state decreased around 50\% within 20~fs according to the dynamics of the ${}^{1}B_{1}-S_0$ states. 
For the dynamics in the ${}^{1}A_{2}-S_0$ states, the excited state population decreases around 10\% within 20~fs.

It is interesting to observe that the off-diagonal density matrix elements $\rho_\mathrm{eg}$ are zero, {which reflects the symmetry property of the non-adiabatic coupling via conical intersection~\cite{Mi127:JPCA23,Neville55:JPB22}.} 
As shown in Fig.~\ref{fig:py_hamil}(c) and (d), the PES of $S_0$ and $^1A_2$ states are invariant under the reflection $Q_{11}\rightarrow-Q_{11}$, but the non-adiabatic coupling term $V_{12}$ is an odd function of $Q_{11}$.
The nuclear wave packet initially in $^1B_1$ state is invariant under the reflection operation, and the wave packet component which is coupled the to $S_0$ state is odd under reflection, so their overlap $\rho_\mathrm{eg}=0$.
For $S_0$ and $^1B_1$ states, the symmetry property under reflection $\theta\rightarrow-\theta$ also causes the off-diagonal matrix elements to be zero.
The opposite parity of the wavepacket is demonstrated in Fig.~\ref{fig:mctdh}, and it is an important property of the conical intersection, namely, the symmetry of the coupling modes.

\subsection{The retrieval of density matrix from ultrafast electron diffraction}
The electronic wavefunction of pyrrole is calculated by the complete active space self-consistent-field (CASSCF) method. 
We use the OpenMolcas package~\cite{Galvan2019:JCTC15} to carry out CASSCF calculation {with the 6-31G(d,p) basis set.}
%
%
An active space of 8 electrons in 7 orbitals for ${}^{1}A_{2}-S_0$ surface coupling and {an active space of 12 electrons in 12 orbitals for} ${}^{1}B_{1}-S_0$ are employed for the CASSCF calculation. {The active space of $^1A_2-S_0$ consists of 3 $\pi$ orbitals, 2 $\pi^*$ orbitals, 1 $\sigma$ orbital and 1 $\sigma^*$ orbital.  In $^1B_1-S_0$ intersections, besides the active orbitals used in $^1A_2-S_0$ calculation, 2 additional occupied $\sigma$ orbitals and 3 virtual $\sigma^*$ orbitals are included in the active space.}
The electron diffraction simulation generates the scattering intensity data with momentum transfer $|\textbf{s}|$ from 0~\AA$^{-1}$ to {2.5}~\AA$^{-1}$. 
For each momentum transfer $\textbf{s}$, 74 Lebedev grid points are used for the rotational averaging. {The numerical convergence of the integration was validated by comparison with 110 Lebedev points with the maximal degree of spherical harmonics of $l_{\mathrm{max}}=17$.}

\begin{figure}
        \includegraphics[width=0.5\textwidth]{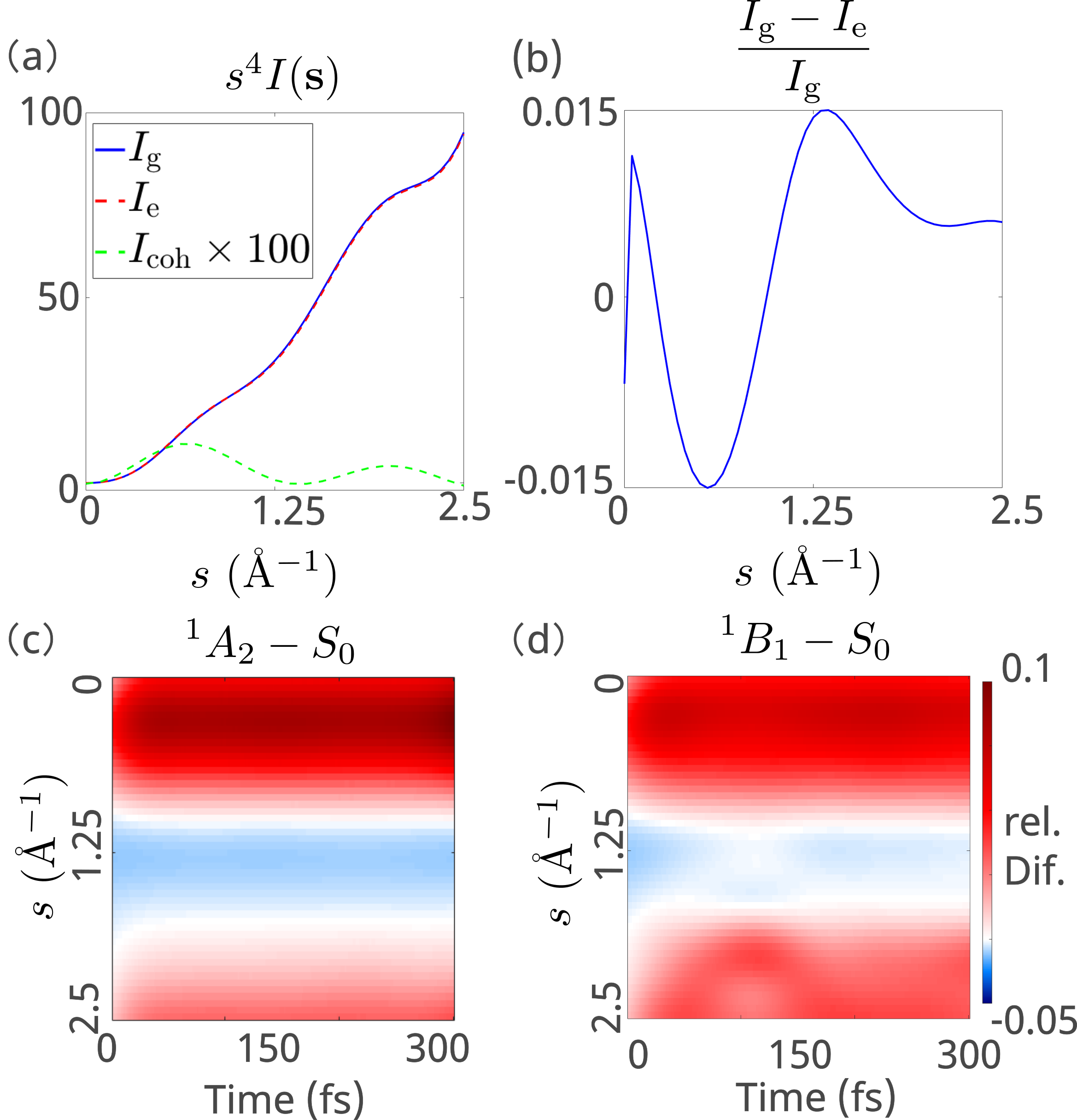}
\caption{\label{fig:ele_cont} Panels {(a)(b)} are the simulated electron diffraction of pyrrole molecule at the ground state equilibrium structure. 
The rotational averaged scattering intensity is plotted for momentum transfer $\textbf{s}$ from 0~\AA$^{-1}$ to 2.5~\AA$^{-1}$. {(a) is the scattering pattern at ground state, excited state, and the coherent part of the
scattering.} {(b)} is the relative difference of scattering intensity between the ground state and excited state. {(c)(d)} are the relative difference change $\dfrac{I_{\mathrm{refine}}(s,t)-I_{\mathrm{refine}}(s,t<0)}{I_{\mathrm{refine}}(s,t<0)}$ from 0~fs to 300~fs in ${}^{1}B_{1}$ and ${}^{1}A_{2}$ state respectively. $I_{\mathrm{refine}}(s,t)$ used for calculation is calculated in Eq.~\ref{eq:Irefine}. $I_{\mathrm{refine}}(s,t<0)$ is the scattering intensity at the ground state. The full width at half maximum (FWHM) of the electron beam is 50~fs.}
\end{figure}

With the ab initio simulation of electron diffraction, we obtain {scattering intensity from the ground
state and the excited state, and the coherent part of the scattering signal}~\cite{Simmermacher2019:PRL122,Kowalewski2017:SD4}.
We present the scattering intensity and contribution at the {ground state equilibrium structure} in Fig.~\ref{fig:ele_cont}. 
Fig.~\ref{fig:ele_cont} shows the temporal evolution of the electron diffraction signal is presented for the photoexcited dynamics of pyrrole initiating from the ${}^{1}B_{1}$ and ${}^{1}A_{2}$ state.  
{The coherent part of scattering contribution
is smaller than the contribution from the ground and excited states, when the momentum
transfer is large.}

\begin{figure}[htb!]
        \includegraphics[width=0.45\textwidth]{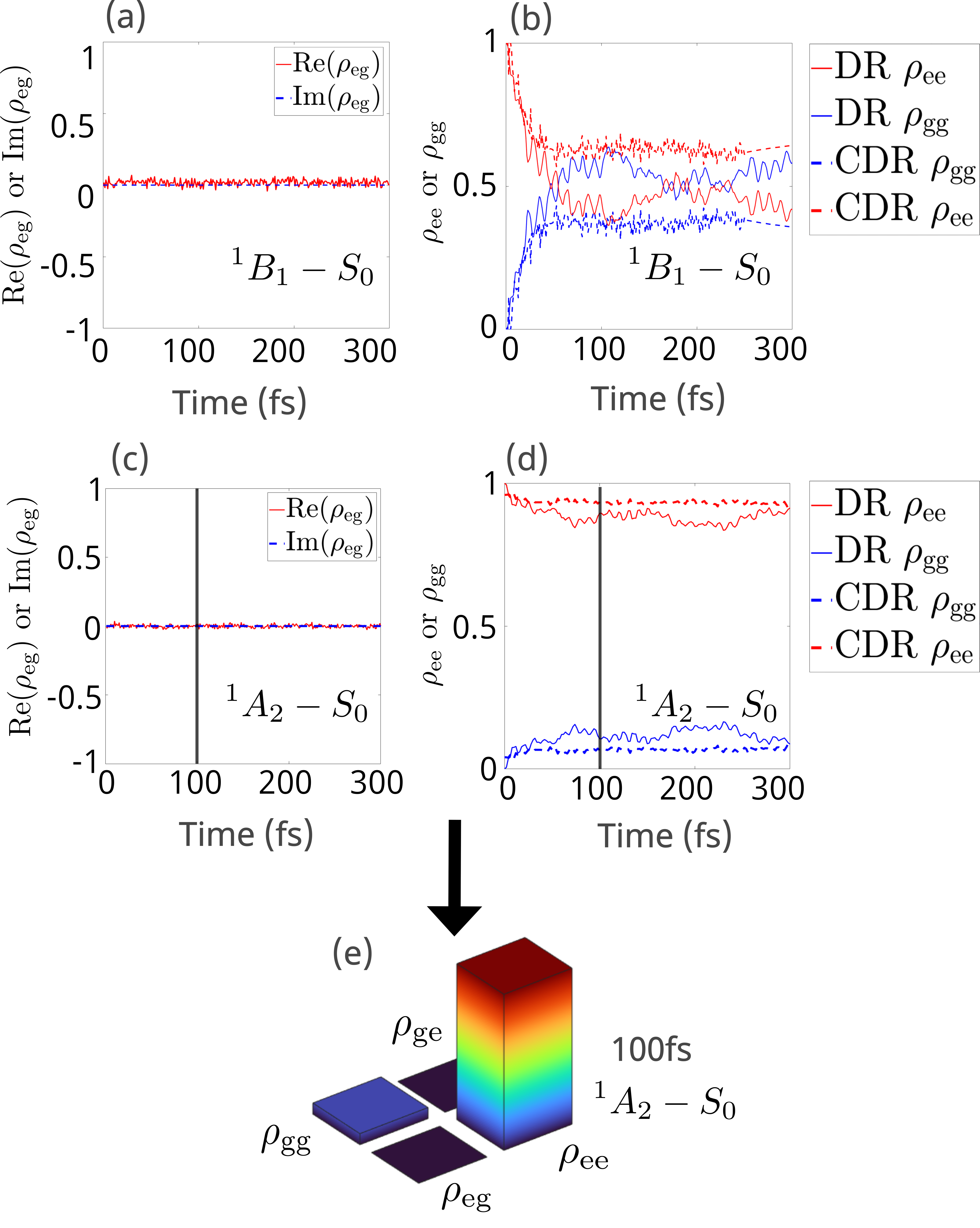}
\caption{ \label{fig:recons} The comparison of {reduced density matrix elements in electronic state basis} between simulated data and the one from electron diffraction data reconstruction.  {(a) is the retrieved real and imaginary part in the non-diagonal density matrix element for the dynamics for the ${}^{1}B_{1}-S_0$ states, (b) is the diagonal terms simulated data and reconstruction comparison in ${}^{1}B_{1}-S_0$ conical intersection.} {(c)} is the retrieved real and imaginary part in the non-diagonal density matrix element for the dynamics for the ${}^{1}A_{2}-S_0$ states, {(d)} is the diagonal terms simulated data and reconstruction comparison in ${}^{1}A_{2}-S_0$ conical intersection.  The "DR" refers to diabatic representation used in MCTDH wavepacket dynamics, "CDR" refers to the crude diabatic representation used for quantum tomography. {(e)} is the reconstruction at 100~fs in ${}^{1}A_{2}-S_0$ dynamics. All the simulation and reconstruction dynamics start from 0~fs and end to 300~fs.}
\end{figure}

To make the simulated data close to the experiment, the left-hand side of Eq.~\ref{eq:total_mat} is substituted for the refined intensity in Eq.~\ref{eq:Irefine}, and 5\% Gaussian noises are added into the data. Using the least square algorithm, we recover the {reduced density matrix element in electronic state basis} $\rho_\mathrm{ee}$ and $\rho_\mathrm{eg}$, and the ground state population $\rho_\mathrm{gg}=1-\rho_\mathrm{ee}$.
The numerical results are shown in Fig.~\ref{fig:recons}. 
Comparing the reconstructed density matrix in the crude diabatic basis with the density matrix in the diabatic basis that is taken from the simulated wavepacket dynamics, it is important to observe that though the two types of density matrices are in slightly different diabatic representations, they are qualitatively consistent in reflecting the population transfer ($\rho_\mathrm{ee}$ and $\rho_\mathrm{gg}$) as well as the interesting symmetry property of the coupling modes and wavepacket ($\rho_\mathrm{eg}$), i.e., the off-diagonal matrix elements are determined to be zero, as expected from the selection rules of the quantum states, {it reveals that $\ket{\chi_\mathrm{e}}$ and $\ket{\chi_\mathrm{g}}$ belong to orthogonal representations of the molecular symmetry group}.
The numerical example presented in this work is a solid demonstration of our method to retrieve the reduced density matrix.
{Further investigation is required for demonstrating the retrieval of density matrix in the case that the off-diagonal density matrix elements are non-zero}.

And most importantly, the reduced density matrices in crude diabatic representation is in practice convenient to obtain, since its reconstruction does not require the knowledge of exact transient nuclear positions in the diabatic PESs, which are physically challenging to determine and would not be even unique in higher dimensional molecular dynamics of polyatomic molecules~\cite{Zhang1998:WS_Theory}.
Thus the crude diabatic density matrix can serve as the suitable candidate for quantum tomography of the reduced density matrix in electronic state basis.

\section{Conclusion}
To summarize, we have demonstrated a quantum tomographic algorithm to reconstruct the {reduced density matrix} in crude diabatic representation from the ultrafast electron diffraction data. 
{The proposed QT algorithm is applied to ultrafast electron diffraction intensities corresponding for the wavepacket dynamics of photoexcited pyrrole molecule, from which we obtain the population evolution of ground state and excited state, as well as the off-diagonal reduced density matrix element reflecting the symmetry properties of electronic and nuclear wavefunctions.}
%
We also demonstrate the advantage and feasibility of using the crude diabatic representation.
{The numerical example of pyrrole wavepacket dynamics, used as a benchmark, illustrates the capability of our QT algorithm to reconstruct the reduced density matrix of other molecular wavepacket dynamics from diffraction data, and its accuracy for molecular dynamics processes without large amplitude atomic motion.
However, it should be checked that the relative change of diffraction intensity caused by atomic motion is small to validate the QT algorithm in this work before applying it to other molecular systems.
}
%
The efficient QT algorithm {paves the way} to make the “molecular movie” at the quantum level {from ultrafast diffraction observables}.

\section*{Acknowledgments}
We thank Oriol Vendrell and Jie Yang for helpful discussions.
JYJ, ASG and RJDM were supported by Natural Sciences and Engineering Research Council of Canada (Grant No.~RGPIN-2019-06518), MZ and ZL acknowledge the support by the National Natural Science Foundation of China (Grant Nos.~12174009, 12234002, 92250303), the National Key R\&D Program (Grant No. 2023YFA1406800), and the Beijing Natural Science Foundation (Grant No.~Z220008).

\section*{AUTHOR DECLARATIONS}

\subsection*{Conflict of Interest}
The authors have no conflicts to disclose.

\section*{Author contributions}
$^*$ J.Y.Z. and M.Z. contribute equally to this work.
\textbf{Jiayang Jiang}: Conceptualization (supporting); Investigation (equal); Writing - original draft (equal); Writing - review \& editing (equal);
\textbf{Ming Zhang}: Conceptualization (supporting); Investigation (equal); Writing - original draft (equal); Writing - review \& editing (equal);
\textbf{Aosheng Gu}: Conceptualization (supporting); Investigation (equal); Writing - original draft (supporting); Writing - review \& editing (supporting);
\textbf{R. J. Dwayne Miller}: Conceptualization (equal); Investigation (equal); Writing - original draft (equal); Writing - review \& editing (equal);
\textbf{Zheng Li}: Conceptualization (equal); Investigation (equal); Writing - original draft (equal); Writing - review \& editing (equal);

\section*{Data Availability}
The data that support the findings of this study are available within the article.

%

\end{document}